\documentclass[letterpaper]{article}
\usepackage{osajnl}
\usepackage[dvips,bookmarks=false]{hyperref}
\begin{document}

\title{Humidity's influence on visible region refractive index structure parameter $C_n^2$}

\author{Mark P. J. L. Chang, Carlos O. Font\raisebox{0.8ex}{\small\it{1,2}},}
\affiliation{\raisebox{0.8ex}{\small\it{1}} Physics Department, University of Puerto Rico, PO Box 9016, Mayag\"uez, Puerto Rico 00681-9016}

\author{G. Charmaine Gilbreath and Eun Oh}
\affiliation{\raisebox{0.8ex}{\small\it{2}} U.S. Naval Research Laboratory, Washington D.C. 20375}

\email{mchang@uprm.edu}

\begin{abstract}
In the near infrared and visible bandpass, optical propagation theory conventionally assumes that humidity does not contribute to the effects of atmospheric turbulence on optical beams.  While this assumption may be reasonable for dry locations, we demonstrate in this paper that there is an unequivocal effect due to the presence of humidity upon the strength of turbulence parameter, $C_n^2$, from data collected in the Chesapeake Bay area over 100-m length horizontal propagation paths.  We describe and apply a novel technique, Hilbert Phase Analysis, to the relative humidity, temperature and $C_n^2$ data to show the contribution of the relevant climate variable to $C_n^2$ as a function of time. 
\end{abstract}


\maketitle 

\section{Introduction}
\label{sect:intro}
It has been known for some time\cite{Roddier1981} that the
scintillation behaviour of point sources is a measure of the optical seeing in
the atmosphere.  What has been less well understood is the contribution of
different environmental variables to optical seeing.  

Over the past decade,
a great deal of study has been dedicated to clarifying this issue.
Comprehensive treatments of the theory of wave propagation in random media are
given in Tatarskii's seminal works\cite{Tatarskii1961,Tatarskii1971}.  Some of the
simplest models based on these complex works are well known and available in
the literature:
Hufnagel--Valley\cite{HufnagelValley1974}, 
SLC-Day and SLC-Night\cite{Miller1976}.
These models are commonly used to predict the strength of weak clear air
turbulence's refractive index structure parameter, $C_n^2$.

The underlying assumption is that the index of refraction of air depends solely upon pressure and temperature at visible and near infrared wavelengths\cite{Tatarskii1961,Roggeman1996}.  We can verify this from landbound horizontal path measurements.  These show that during the daylight hours on a clear sunny day, the weak scintillation r\'{e}gime $C_n^2$ trend is dominated by solar insolation, as is illustrated by the left hand plot of Fig. \ref{fig:solarinsolation}.

The right hand plot of Fig. \ref{fig:solarinsolation} illustrates the well known fact that the effects of turbulence do not cease after sunset.  Here it is less clear as to the predominant contributing factors.  The assumption of temperature dominance is sensible in the driest of environments, but we will demonstrate from experimental evidence that this is an incomplete condition where there is a significant amount of water vapour entering or leaving air volume where $C_n^2$ is measured.

\subsection{Past work}

A body of work has been reported whose aims have been to compare bulk climate estimates of $C_n^2$ with optically derived readings\cite{Frederickson1999,Davidson1981,Kunz1996}.  In those works, the wavelengths of the optical measurements were in the mid--infrared (3 to 5 $\mu$m), and all showed that humidity along the measurement volume plays a significant r$\hat{o}$le in the temporal behaviour of $C_n^2$.

In a theoretical study of the structure parameter using bulk measureables over sea ice and snow, Andreas \cite{Andreas1988} defined a refractive index scale $n_*$ via
\begin{equation}
\label{eqn:n*}
n_* = A t_* + B q_* \rightarrow \frac{n_*}{A t_*}  =  1 + \frac{B q_*}{A t_*}
\end{equation}
where $t_*$ and $q_*$ represent temperature and humidity, scaled such they are constant with height and $A$ and $B$ are constants given in Appendix A.  The variable $n_*$ therefore represents a horizontal refractive index layer above the terrestrial surface.  He rewrote Eq. (\ref{eqn:n*}) in terms of the Bowen ratio $Bo$, the ratio of sensible heat flux to latent heat flux, such that
\begin{equation}
\label{eqn:n*/At*}
\frac{n_*}{A t_*} = 1 + \left( \frac{\rho c_p}{L} \right) \frac{B}{A (Bo)} = 1 + \left( \frac{B}{K A (Bo)} \right)
\end{equation}
where $c_p$ represents the specific heat of air at constant pressure, $L$ the latent heat of sublimation of ice and $\rho$ is the air density.  Since $Bo \propto t_* / q_*$, this formulation allows the dependence of the refractive index scale upon temperature and humidity scales to be easily interpreted.  

We adapt Eq. (\ref{eqn:n*/At*}) to our experimental conditions by setting $L$ to be the latent heat of vapourisation of water, and we assume $P = 1000$ hPa, $T = 25$ $^\circ$C and $\lambda = 0.93$ $\mu$m. The result is shown in Fig. \ref{fig:n*_vs_Bo}; from this it is clear that when $|Bo|$ is large, $n_*/A t_*$ is near 1.  This means that when sensible heat (heat exchange without change of thermodynamic phase) dominates the Bowen ratio, the $t_*$ in the denominator of Eq. (\ref{eqn:n*}) is the principal parameter.

When $|Bo|$ is small however, the latent heat flux dominates so $q_*$ makes the major contribution to $n_*$.  Thus there is an interplay between humidity and temperature effects upon the refractive index in the visible/near infrared, which vary as a function of the Bowen ratio.  The pole at the origin of Fig. \ref{fig:n*_vs_Bo} is indicative of a breakdown of this formulation at very small Bowen ratios.  Nevertheless the curve strongly suggests that as long as there is significant injection of moisture into the measurement volume of $C_n^2$, humidity effects become non-negligible.

In this paper we demonstrate that the temporal behaviour of the optically determined turbulence structure parameter $C_n^2$ in the visible and near infrared bandpass at a coastal site is strongly influenced by local humidity.  Although no Bowen ratio values could be determined from the available measurements, the results do support the qualitative interpretation of Andreas' work.

\section{Experiment and Correlogram Analysis} 
\label{sect:expt+corr}

The $C_n^2$ and associated weather variable data used in this study were collected over a number of days during 2003 and 2004 at Chesapeake Beach in Maryland, USA.

The $C_n^2$ data were obtained with a commercially available scintillometer (model LOA-004) from Optical Scientific Inc, which serves as both a scintillometer and as an optical anemometer for winds transverse to the beam paths.  The separation between source and receiver was just under 100-m, placing the system in the weak scintillation r\'{e}gime.  The local weather parameters were determined by a Davis Provantage Plus (DP+) weather station.  The LOA-004 had a sample rate of 10 seconds, while the DP+ was set at 5 minutes.

The LOA-004 instrument comprises of a single modulated infrared transmitter whose output is detected by two single pixel detectors.  The transmitting LED has a bandwidth of 0.92 to 0.96 $\mu$m, while the detector bandwidth is much broader, at $0.65$ to $1.0$ $\mu$m.  The path integrated $C_n^2$ measurements are determined by the LOA instrument
by computation from the log--amplitude scintillation ($C_\chi(r)$) of the two receiving signals\cite{Ochs1979,Wang2002}.  The algorithm for relating $C_\chi(r)$ to $C_n^2$ is based on an equation for the log--amplitude covariance function in Kolmogorov turbulence by Clifford {\em{et al.}}\cite{Clifford1974}.

The $C_n^2$ data was smoothed with a 60 point (5 minute) rolling average function.  We define the morning and night portions of a 24 hour period with respect to the measured solar irradiance function, such that we exclude the effect of solar insolation from the data in this study.  Morning runs from midnight until sunrise, while night runs from sunset until 23:59.  As reported in Oh {\em et al.}\cite{Oh2004a,Oh2004b,Oh2004c} visual inspection of the time series data gives the impression that there is an approximate inverse relationship between $C_n^2$ and relative humidity.  This can be appreciated in a more quantitative manner by graphing one parameter against the other.

We chose data sets in which the temperature variations are no more than $16^\circ$ F  and the pressure change is at most 15 hPa over the time intervals of interest.  The actual range of variation per data set is shown in table \ref{tbl:RHTPranges}.  The data sections were also selected to have no scattering effects due to snow or rain, and the wind was northerly (to within approximately $\pm 20^\circ$, inflowing from the bay to land).
Only a small subset eight morning and evening runs, spanning seven days between November 2003 and March 2004, provided complete time series in both ambient weather variables and $C_n^2$. Part of this is shown in Fig. \ref{fig:Mar2Apr}; the $C_n^2$ vs. humidity correlograms evidence a negative gradient in all eight runs.  Table \ref{tbl:RHTPranges} shows us that the relative humidity variation is more strongly reflective of changes in absolute moisture content than of moisture holding capacity of the air.  

We conclude that humidity plays a significant part in the behaviour of the refractive index structure parameter $C_n^2$ in the datasets studied.

\section{The Stationarity Problem}
\label{sect:stationarity}

The unsophisticated use of the correlogram tool in Section \ref{sect:expt+corr} is a rapid, first order method for inferring the statistical influence of one measurable upon another.  However it is unsatisfactory as it does not reveal any detailed information, such as exactly when the humidity contribution is important with respect to all other possible parameters (e.g. temperature) and to what degree its contribution is influential.  Cross covariance techniques are a possible second order method\cite{Font2006a} for extending the analysis, but there are two major stumbling blocks.  The first is the difficulty in interpreting the meaning of the covariance functions physically and the second is the non--stationary nature of the data.  In the latter case it is well known that strict stationarity is a constraint that is impossible to satisfy on practical grounds, since no detector can cover all possible points in phase space.  This has motivated us to employ a novel analysis technique to infer the relationship between $C_n^2$ and climate parameters, which we briefly describe in the following section.  We leave a more detailed study of the limitations of the technique to a separate paper.

\subsection{The Analytic Signal}
Gabor\cite{BornWolf1999} defined the complex analytic signal, namely
\begin{eqnarray}
\label{eqn:Gabor1}
\Psi (t) & = & X(t) + i Y(t) \\ \nonumber
\textrm{where} ~~ Y(t) & = & {\cal{H}}[ X(t) ] = \frac{-1}{\pi} {\bf{P}} \int_{-\infty}^{\infty} \frac{X(\tau)}{(t - \tau)} d\tau
\end{eqnarray}
where ${\cal{H}} [ \bullet ]$ represents a Hilbert Transform.  
Through this definition, $\Psi(t)$ is unique.  The Hilbert Transform is a well known integral transform with a singular kernel $(1/(\pi (t - \tau))$, $\tau$ also being a time variable if $t$ is time.  It is also a Cauchy Principal integral, which we denote by ${\bf{P}}$.  Note that the Hilbert Transform preserves the norm of the real signal; the difference between $Y$ and $X$ is a $\pi / 2$ phase shift for a periodic function.  That is, the Hilbert Transform of $\cos t$ is $\sin t$, and $\sin t$ is $- \cos t$, and the transform of a constant is zero.

Eq. (\ref{eqn:Gabor1}) means that we can write the analytic signal as
\begin{eqnarray}
\label{eqn:Gabor2}
\Psi(t) = a(t) \exp^{i \Phi(t)} ~,~~~\textrm{where} ~~~ a(t) & = & \sqrt{X^2 (t) + Y^2 (t)} \\ \nonumber
\Phi(t) & = & \arctan \left( \frac{Y(t)}{X(t)} \right)
\end{eqnarray}
which is similar to the well known Fourier expression.  We may now determine the instantaneous signal phase, $\Phi(t)$.
Thus we can also calculate the instantaneous frequency $\omega(t)$, defined as
\begin{equation}
\omega (t) = \frac{d \Phi (t)}{dt}.
\end{equation}
Both $\Phi(t)$ and $\omega(t)$ can be interpreted as physical measureables, provided certain preconditions are met which we describe below.

\subsection{Hilbert Phase Analysis}
There is a problem with the definition of Eq. (\ref{eqn:Gabor1}); the Hilbert Transform's kernel represents a non--causal filter, of infinite support.  So if one applies the Hilbert Transform directly to a time varying signal which has a non-zero local mean in any subsection, there is a high probability that at least one of a number of paradoxes{\cite{Cohen1995}} will be encountered.  This may be best appreciated if we consider a signal in phase space, as in the left hand plot of Fig. \ref{fig:hilbertphasespace}. Here we show the real versus imaginary components of the Hilbert Transform of a series of analytic $C_n^2$ measurements.  The trajectory of the analytic signal's vector is subject to many alterations in the vector start point, norm and phase angle.  Any attempt to determine the instantaneous frequency is bound to be problematic as we are not able to follow the signal vector's start position over time; this starting point obviously does not remain at the origin of the coordinate system.  Calculating the instantaneous frequency from the Hilbert Transform as is will generate both positive and negative values, rendering it physically uninterpretable.

The paradoxes may be avoided by the application of the Empirical Mode Decomposition (EMD) method developed by Huang {\em{et al}}\cite{Huang1998}, which we have implemented{\cite{Chang2006, Font2006b}}.  EMD is a unique and novel method that is able to separate an arbitrary real time series into ``eigenfunctions'' termed Intrinsic Mode Functions or IMFs, each of which possesses a structure with well defined instantaneous frequencies, $\omega(t)$.  The term eigenfunction is used suggestively here; we do not mean to imply that the IMFs are eigenfunctions in the strict sense.

Briefly, the EMD technique consists of (1) the determination of two envelope functions about the time series, covering all the local minima and maxima respectively; (2) the computation of the mean of the two envelope functions; (3) the removal of the mean of the envelopes from the time series in an iterative manner, until the mean is found to be zero; (4) the storing of the resulting mode as an IMF.  The IMF is then subtracted from the original time series and steps (1) to (4) are repeated, thereby sifting out a family of IMFs, stopping only when the resulting mode shows no variation.  The final mode represents the overall trend of the signal and is not itself an IMF.

The Hilbert Transform of the IMFs, one of which is shown on the right side of Fig. \ref{fig:hilbertphasespace}, ensures that the analytic signal vector's origin stays fixed and no sudden changes in the direction of $\omega (t)$ occur.  These conditions being satisfied, the instantaneous frequency remains positive and physically meaningful.

\section{Hilbert Phase Analysis}
\label{sect:HPA}

The {\em Hilbert Phase Analysis} (HPA) technique is based on the ideas mentioned in the previous subsections.  It is clear that a phase angle ($\Phi(t)$), as well as an amplitude ($a(t)$), can be found from the Hilbert Transform of the IMFs derived from the EMD sifting process.

The procedure for the HPA data analysis follows a three step protocol:
\begin{enumerate}
\item Decompose the time series measurements of differing parameters via EMD, obtaining their IMFs and trend lines.
\item Apply the Hilbert Transform to the various IMF sets.  We discard the trend lines as the reason that the original time series fail to be Hilbert Transformed in a physically comprehensible manner.
\item Examine the instantaneous phase angles of the Hilbert Transformed IMFs between different parameters to infer the dynamics of the physical system.
\end{enumerate}

\subsection{Interpretation of HPA}

In this section we demonstrate that physical effects of a non--linear, non--stationary, time varying system can be studied via the sum of all IMF phases, $\sum_{IMF} \Phi(t)$.

In Fig. \ref{fig:solarhpa} the $\sum_{IMF} \Phi(t)$ is graphed with the solar radiation data superimposed for example days.  There is clearly a gradient change in the phase function at sunrise and sunset.  Also notable is the phase jump towards a lower gradient whenever the solar radiation function exhibits a drop in amplitude; likewise the phase gradient increases with sudden increases in the measured solar radiation.  This can be understood as: a reduction in energy in the system leads to a lower instantaneous frequency, therefore we see a lower phase gradient.  A change in energy results in a change in instantaneous frequency, so we see a modified gradient.

We conclude that overall physical features of a non--stationary time series can be extracted by inspection of the sum of its Hilbert Phases of the IMFs.

\section{HPA of $C_n^2$, humidity and temperature}
\label{sect:cn2humtemp}

\subsection{Phase locking between measureables}
To better understand the dependence of $\Phi_C$ upon humidity and temperature, we consider the difference between observable phases, as illustrated in Fig. \ref{fig:DeltaPhi11102003morning}.  We define the following difference terms,
\begin{eqnarray}
\Delta_{CT} & = & \sum \Phi_C - \sum \Phi_T ~,~~~
\Delta_{CH} = \sum \Phi_C - \sum \Phi_H \\ \nonumber
\Delta_{C\overline{HT}} & = & \sum \Phi_C - \frac{(\sum \Phi_H + \sum \Phi_T)}{2}
\end{eqnarray}
representing the phase differences between $C_n^2$ and the bulk climate parameter.  It is anticipated that the controlling climate parameter will be indicated by a near zero difference, $\Delta$.  We reason this because if $\Delta$ is constant between $C_n^2$ and a climate variable then the two datasets must be phase locked.  If the mean $\Delta$ is zero, then any variations about zero should indicate a synchronization between the datasets.  Supporting empirical evidence is found upon studying cases wherein the solar insolation influence upon the measured $C_n^2$ is strong, as in Fig. \ref{fig:HPAevidence}.  In those plots one can see that the difference between the summed Hilbert Phase of $C_n^2$ and solar radiation ($\sum \Phi_C - \sum \Phi_S$) flattens out when the solar function is significant.

Motivated by this, we determined by linear regression the mean gradients of all the $\Delta$ curves, which we list in Table \ref{tbl:deltagradients}.  If we assume that the only major contributor should be the local temperature variations, $\Delta_{CT}$ should be zero or near zero in all cases.  This turns out not to have occurred.  In fact, $\Delta_{CT}$ has no near zero value, so temperature variation cannot be said to be phase locked with the refractive index structure parameter.

If humidity has a controlling effect in the visible / near infrared, then it should exhibit a range of values near zero for $\Delta_{CH}$.  It only does so for 11/10/03 (am) and 03/28/04 (am).  It is noteworthy that 11/09/03 shows a possible phase lock between the mean of humidity and temperature with $C_n^2$.  This seems to indicate that that temperature is vying with humidity for influence over the $C_n^2$ parameter during that morning.

\subsection{Further study}

If we define ``dominance'' to mean that the phase gradient is the value closest to zero of the differences under scrutiny then we find that humidity is dominant for the 11/10/03, 02/02/04, 03/27/04 and 03/28/04 datasets.  The 11/03/03 and 04/03/04 (a.m. and p.m.) datasets show that $\Delta_{CT} < \Delta_{CH}$, indicating the dominant effect is temperature, this being extremely strong for the 04/03/04 (pm) dataset and rather weaker for the other two.  We postulate that the 11/03/03 and 04/03/04 datasets indicate both temperature and humidity are contributing to the behaviour of $C_n^2$, with the HPA method possibly showing the proportional contribution to $C_n^2$ of each climate variable.  This is clearly an area to be examined further.

\section{Conclusions}
\label{sect:conclusions}

From experimental data we have shown conclusively that humidity plays a significant part in the visible/near infrared measure of $C_n^2$ in a coastal environment.  This is in qualitative agreement with Andreas' model and is a natural extension of the results from the mid infrared.

Furthermore we have explored a new technique, Hilbert Phase Analysis, with which to study this physical phenomenon.  
In overall terms, the HPA method is in agreement with the correlogram results.  Phase locking between data is an unexpected result that needs further examination.  We have found that the HPA technique described here is very promising and is likely to provide much more information about the changes to a physical system than traditional methods.

\appendix

\section*{Appendix A: Definitions}
\setcounter{equation}{0}
\renewcommand{\theequation}{A{\arabic{equation}}}

The constants $A$ and $B$ of Eq. \ref{eqn:n*} in the 0.36 to 3 $\mu$m wavelength region are
\begin{eqnarray}
A & = & -10^{-6} m_1(\lambda) (P/T^2) \\ \nonumber
B & = & 4.6150 \times -10^{-6} [m_2(\lambda) - m_1(\lambda)].
\end{eqnarray}
$P$ and $T$ are the ensemble average air pressure and temperature, respectively.  The functions $m_1$ and $m_2$ are polynomials in wavenumber given by
\begin{eqnarray}
m_1(\lambda) & = & 23.7134 + \frac{6839.397}{130 - (1/\lambda)^2} + \frac{45.473}{38.9 - (1/\lambda)^2} \\ \nonumber
m_2(\lambda) & = & 64.8731 + 0.58058 (1/\lambda)^2 - 0.0071150 (1/\lambda)^4 + 0.0008851 (1/\lambda)^6
\end{eqnarray}
with $\lambda$'s units in micrometres.  

\section{Acknowledgements}
MPJLC is grateful to Norden Huang, Haedeh Nazari and Erick Roura for illuminating discussions.
Part of this work was supported by the Office of Naval Research.

\bibliography{HumCn2bib}
\bibliographystyle{osajnl}


\newpage

\section*{List of Figure Captions}

\noindent Fig. 1. The effects of solar insolation on $C_n^2$ as measured on March 9, 2006 in Puerto Rico.  Solar radiation measurements are superimposed at an arbitrary scale on top of the $C_n^2$ data.  (Left) The full 24 hour period.  The vertical axis values are of order $10^{-12}$ $m^{-2/3}$. (Right) Close up of the evening period.  The vertical axis values are of order $10^{-14}$ m$^{-2/3}$.  The $C_n^2$ data were obtained at an urban site 1.75-km from the sea.  These values are in agreement with measurements of seeing over the sea by alternative means\cite{Beaumont1997}.

\noindent Fig. 2. Graph of $n_*/(A t_*)$ vs Bowen ratio for $\lambda = 0.93$ $\mu$m.

\noindent Fig. 3. Example correlograms of $C_n^2$ and Relative Humidity in the absence of solar insolation.  The upper and lower bounds indicate the 50\% confidence level.  The $C_n^2$ magnitudes are $10^{-15}$ $m^{-2/3}$, in agreement with measurements of seeing over the sea by alternative means\cite{Beaumont1997}.

\begin{figure}[htbp]
\centering
\includegraphics{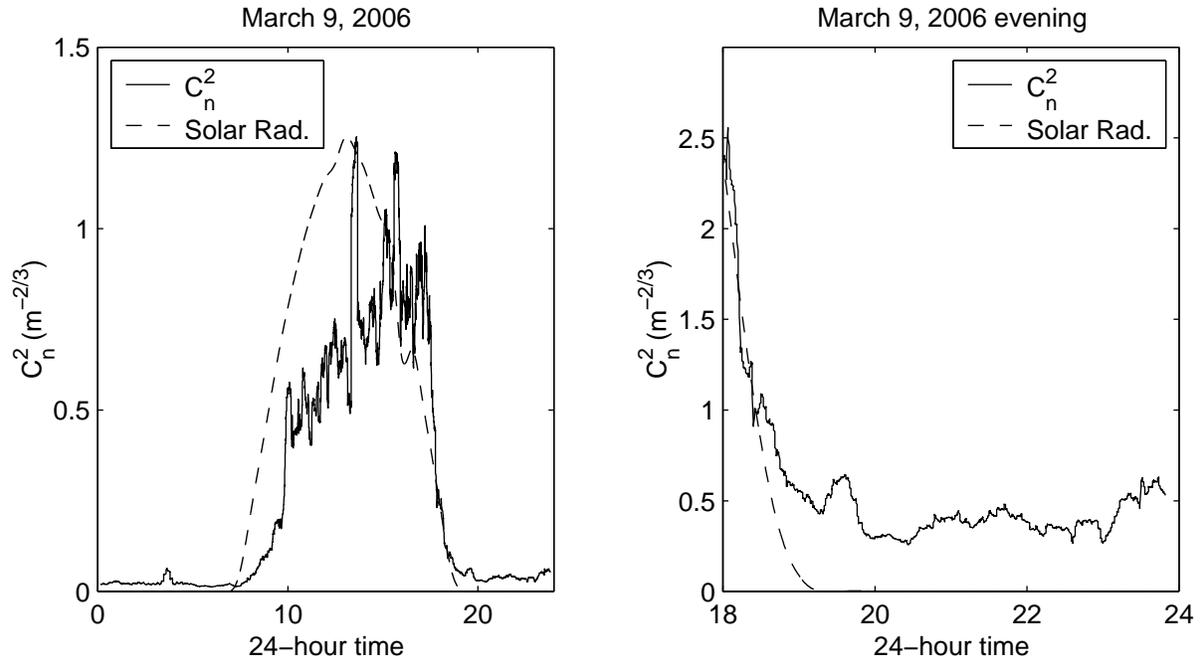}
\caption{\label{fig:solarinsolation} The effects of solar insolation on $C_n^2$ as measured on March 9, 2006 in Puerto Rico.  Solar radiation measurements are superimposed at an arbitrary scale on top of the $C_n^2$ data.  (Left) The full 24 hour period.  The vertical axis values are of order $10^{-12}$ $m^{-2/3}$. (Right) Close up of the evening period.  The vertical axis values are of order $10^{-14}$ m$^{-2/3}$.  The $C_n^2$ data were obtained at an urban site 1.75-km from the sea.}
\end{figure}

\begin{figure}[htbp]
\centering
\includegraphics{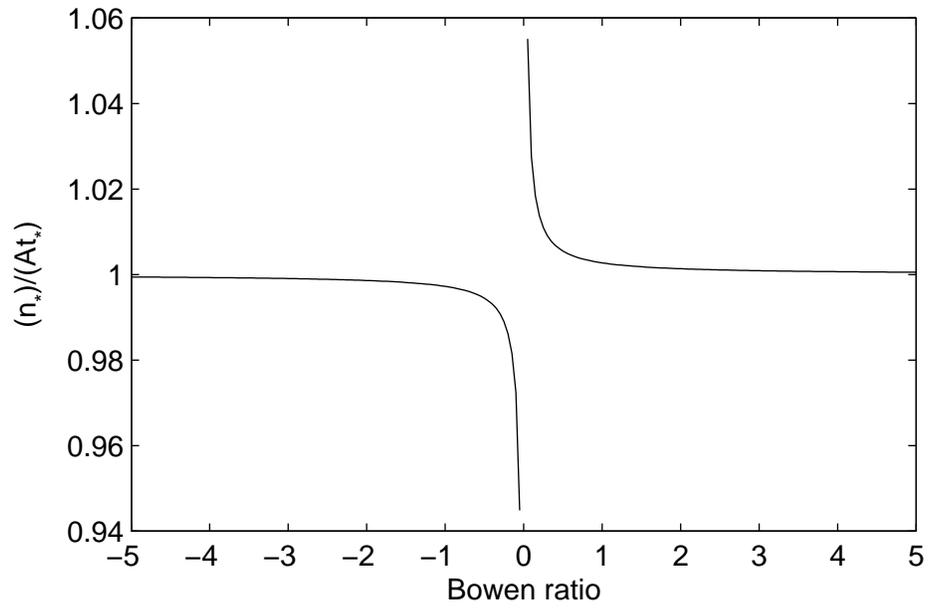}
\caption{\label{fig:n*_vs_Bo} Graph of $n_*/(A t_*)$ vs Bowen ration for $\lambda = 0.93$ $\mu$m.}
\end{figure}

\begin{figure}[htbp]
\centering
\includegraphics{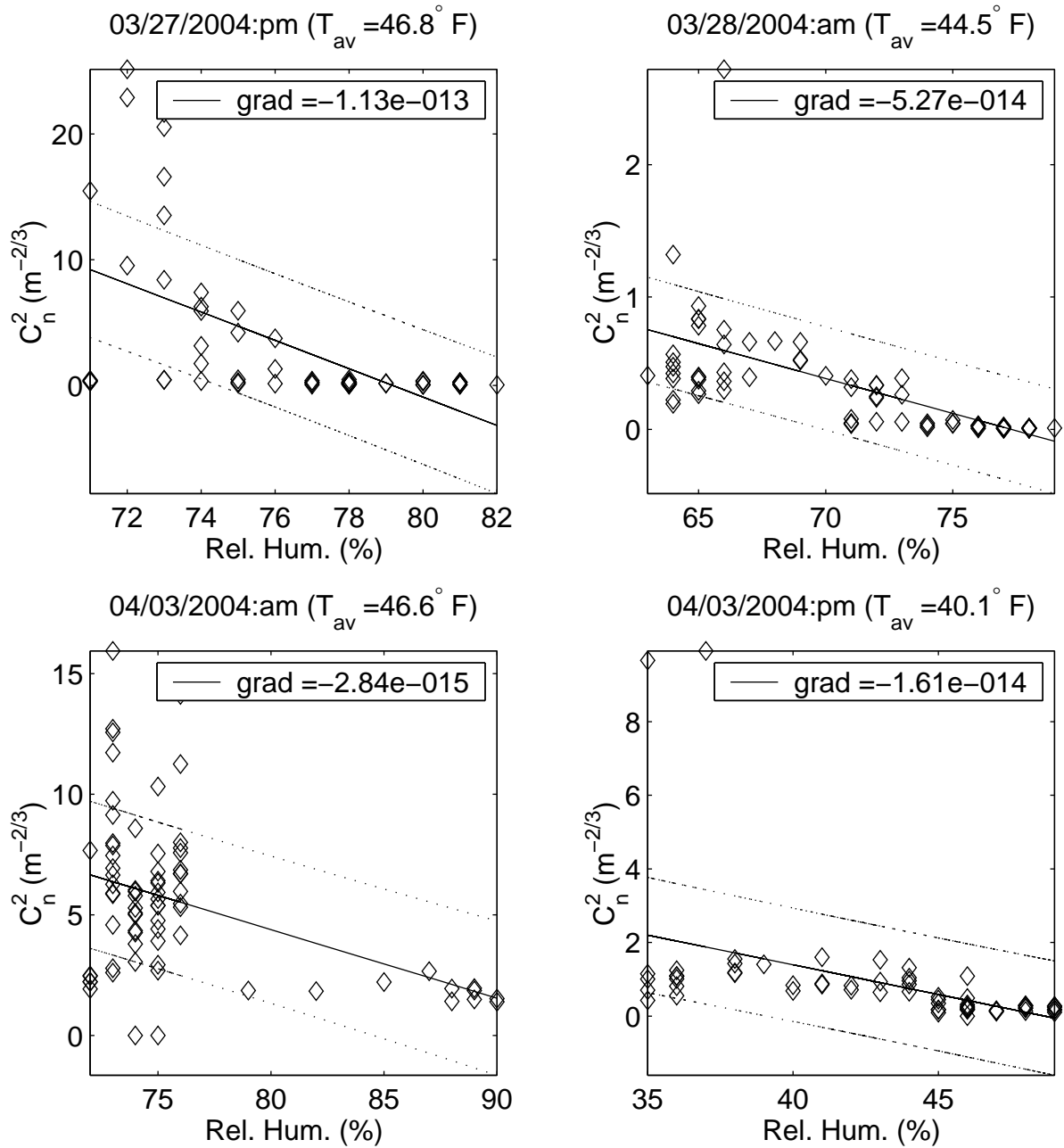}
\caption{\label{fig:Mar2Apr} Example correlograms of $C_n^2$ and Relative Humidity in the absence of solar insolation.  The upper and lower bounds indicate the 50\% confidence level.  The $C_n^2$ magnitudes are $10^{-15}$ $m^{-2/3}$, in agreement with measurements of seeing over the sea by alternative means\cite{Beaumont1997}.}
\end{figure}

\begin{figure}[!htp]
\begin{center}
\begin{tabular}{c}
\includegraphics{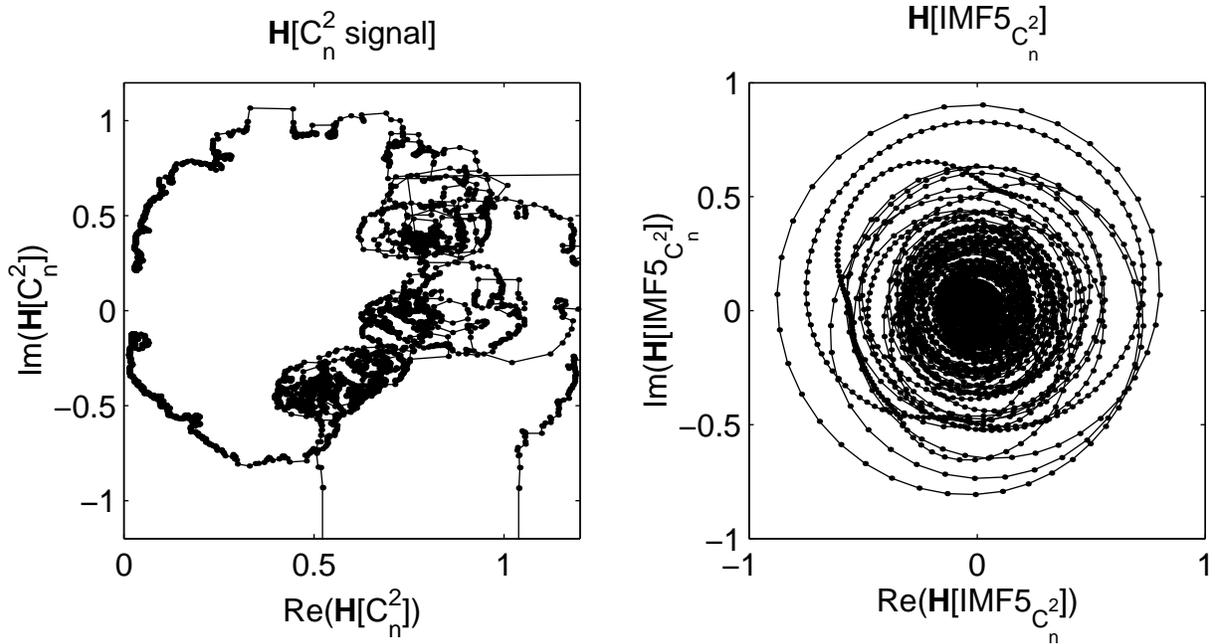}
\end{tabular}
\end{center}
\caption{\label{fig:hilbertphasespace}(left) Hilbert Phase Space plot of the trajectory of the $C_n^2$ signal vector of figure \ref{fig:solarinsolation}. The signal vector's start point drifts arbitrarily around the phase space, making the determination of a physically reasonable instantaneous frequency impossible over the whole path.  (right) Hilbert Phase Space plot of the trajectory of a single IMF derived from the $C_n^2$ signal.  The IMF vector's start point is stable and its trajectory does not change direction, so a positive instantaneous frequency can be determined at all points.}
\end{figure}

\begin{figure}[htbp]
\centering
\includegraphics{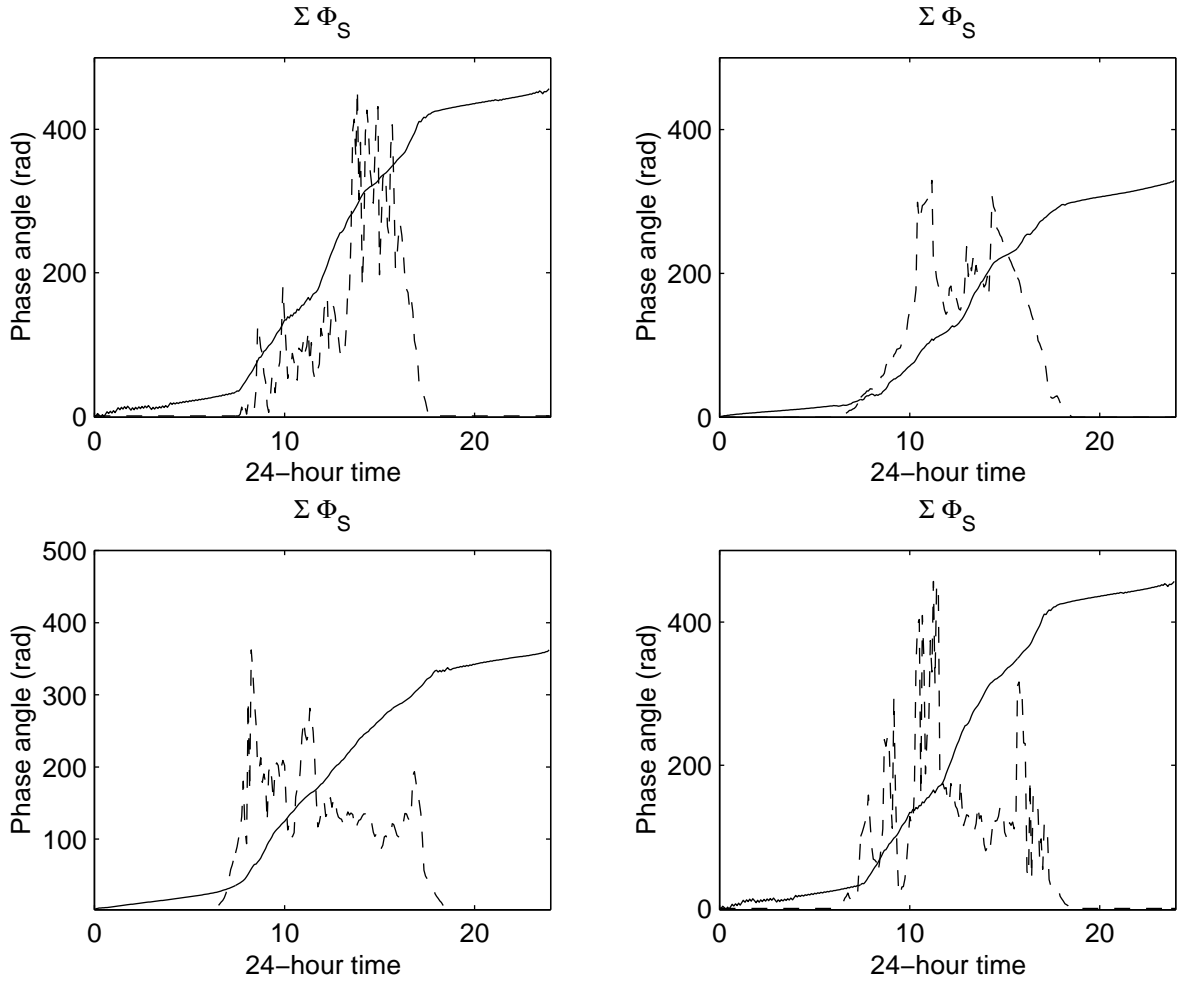}
\caption{\label{fig:solarhpa} Sum of all Hilbert Phases of the measured Solar Radiation IMFs ($\sum \Phi_S$) for the following days: Feb 2, Mar 27, Mar 28 and Apr 3, 2004.  The Solar Radiation is superimposed at an arbitrary scale.}
\end{figure}

\begin{figure}[htbp]
\centering
\includegraphics{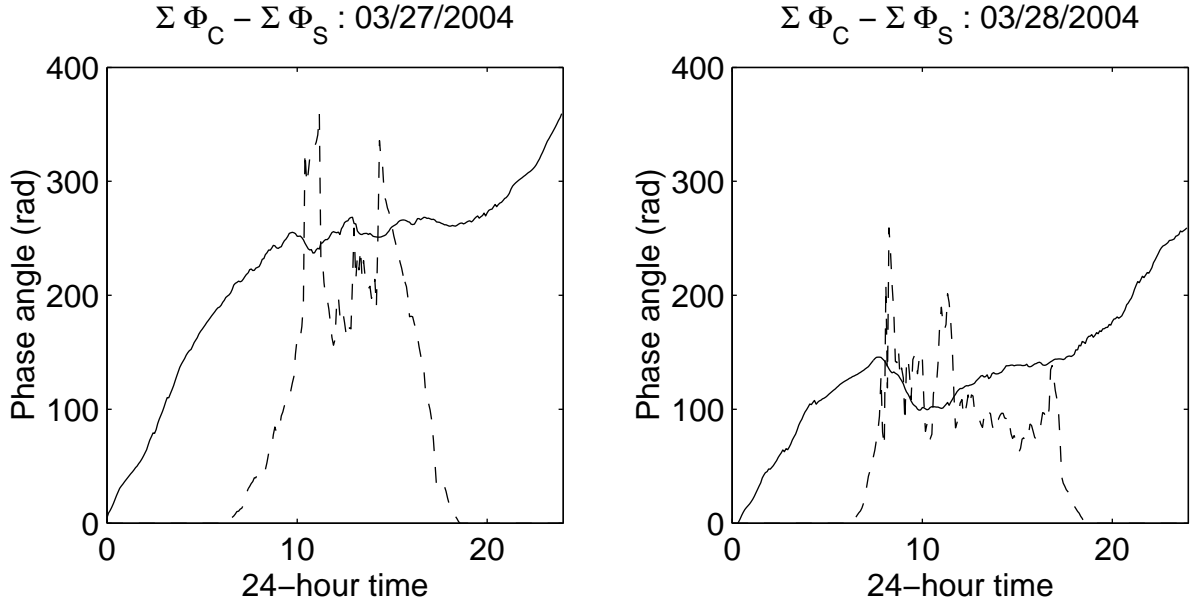}
\caption{\label{fig:HPAevidence} Plots of Hilbert Phase differences between $C_n^2$ and solar radiation for 03/27/04 and 03/28/04.  The solar radiation function is superimposed at an arbitrary scale on each graph.  Note the flattening out of the phase difference function during the daylight hours.}
\end{figure}

\begin{figure}[htbp]
\centering
\includegraphics{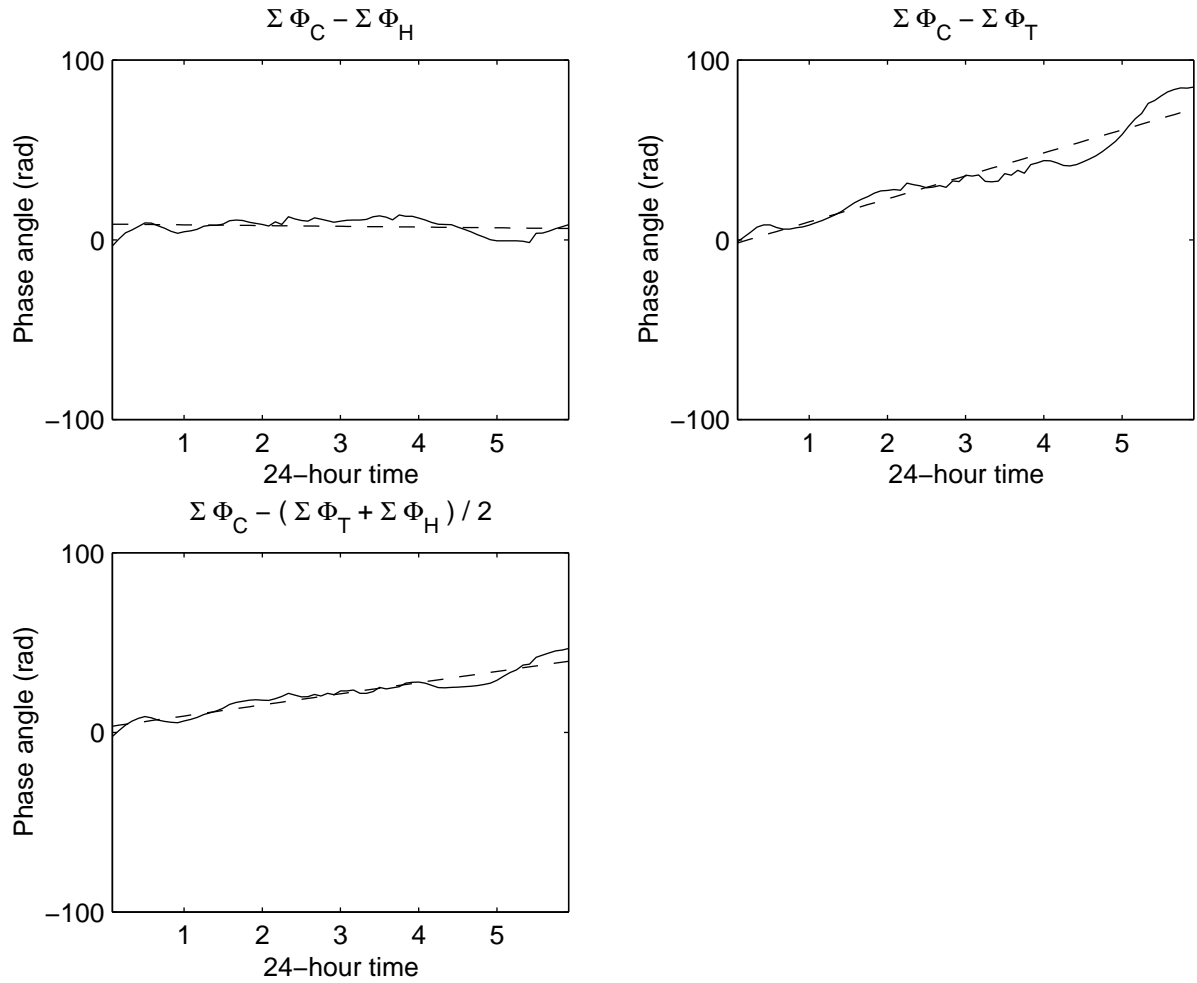}
\caption{\label{fig:DeltaPhi11102003morning} Difference plots for the morning of November 10, 2003.  The dotted line is a linear regression, estimating the mean phase gradient.  The top left graph shows a phase lock between $C_n^2$ and relative humidity.}
\end{figure}

\noindent Fig. 4. (left) Hilbert Phase Space plot of the trajectory of the $C_n^2$ signal vector of Fig. \ref{fig:solarinsolation}. The signal vector's start point drifts arbitrarily around the phase space, making the determination of a physically reasonable instantaneous frequency impossible over the whole path.  (right) Hilbert Phase Space plot of the trajectory of a single IMF derived from the $C_n^2$ signal.  The IMF vector's start point is stable and its trajectory does not change direction, so a positive instantaneous frequency can be determined at all points.

\noindent Fig. 5. Sum of all Hilbert Phases of the measured Solar Radiation IMFs ($\sum \Phi_S$) for the following days: Feb 2, Mar 27, Mar 28 and Apr 3, 2004.  The Solar Radiation is superimposed at an arbitrary scale.

\noindent Fig. 6. Plots of Hilbert Phase differences between $C_n^2$ and solar radiation for 03/27/04 and 03/28/04.  The solar radiation function is superimposed at an arbitrary scale on each graph.  Note the flattening out of the phase difference function during the daylight hours.

\noindent Fig. 7. Difference plots for the morning of November 10, 2003.  The dotted line is a linear regression, estimating the mean phase gradient.  The top left graph shows a phase lock between $C_n^2$ and relative humidity.

\noindent Table 1.  Mean and range of bulk parameters measurements and the maximum range of specific humidity ($\Delta_{Q_s}$).

\noindent Table 2. Linear regression line gradients of the Phase Differences.

\newpage

\newpage

\newpage

\newpage

\newpage

\newpage

\newpage

\newpage

\begin{table}[htbp]
\caption{ \label{tbl:RHTPranges} Mean and range of bulk parameters measurements and the maximum range of specific humidity ($\Delta_{Q_s}$).}
\centering
\begin{tabular}{cccccccc} \hline
Date 		& $\overline{RH}$
		& $\overline{T}$
		& $\overline{P}$
		& $\Delta_{RH}$
		& $\Delta_{T}$
		& $\Delta_{P}$
		& $\Delta_{Q_s}$
\\
		& (\%)
		& ($^\circ$F)
		& (hPa)
		&
		&
		&
		& (g/kg)
\\ \hline
11/03/03 (pm) 	& 83.6			
		& 65.2			
		& 1015.3			
		& 15	
		& 2
		& 0.4 
		& 2.2
\\
11/09/03 (am)	& 49.3
		& 38.8
		& 1030.9
		& 11
		& 3.1
		& 1.2
		& 0.5
\\
11/10/03 (am)	& 76.4	
		& 46.8	
 		& 1027.6	
		& 11
		& 8
		& 1.8
		& 0.6
\\
02/02/04 (am)	& 91.4
		& 29.9
		& 1022.9	
		& 16
		& 2.5	
		& 6.6
		& 0.8
\\
03/27/04 (pm)	& 76.4	
		& 46.8	
		& 1027.6	
		& 11 
		& 8 
		& 1.8
		& 1.7
\\
03/28/04 (am)	& 71	
		& 44.5	
		& 1027.6
		& 16
		& 0.7	
		& 0.6
		& 0.9
\\
04/03/04 (am)	& 76.1
		& 46.6	
		& 997.4	
		& 18
		& 5.9 	
		& 4.9
		& 0.4
\\
04/03/04 (pm)	& 43	
		& 40.1	
		& 1004.5
		& 14
		& 5.6
		& 2.1
		& 1.1
\\ \hline
\end{tabular}
\end{table}

\newpage

\begin{table}[htbp]
\caption{ \label{tbl:deltagradients} Linear regression line gradients of the Phase Differences.}
\centering
\begin{tabular}{cccc} \hline
Date 			&$\Delta_{CH}$ 	& $\Delta_{CT}$ 	& $\Delta_{C \overline{HT}}$ 	\\ \hline
11/03/03 (pm)		& -9.7(4)	& -8.0(0)		& -8.8(7)			\\
11/09/03 (am)		& -11.2(4)	& 9.7(7)		& -0.7(4)			\\
11/10/03 (am)		& -0.4(1)	& 12.7(9)		& 6.1(9)			\\
02/02/04 (am)		& 8.5(1)	& -12.3(0)		& -1.8(9)			\\
03/27/04 (pm)		& 3.7(6)	& 28.1(7)		& 15.9(7)			\\
03/28/04 (am)		& 0.0(2)	& -3.3(1)		& -1.6(5)			\\
04/03/04 (am)		& 6.8(1)	& 6.6(8)		& 6.7(4)			\\
04/03/04 (pm)		& -27.3(3)	& -2.3(3)		& -14.8(3)			\\ \hline
\end{tabular}
\end{table}

\end{document}